\documentclass[aps,accepted=2017-07-10]{quantumarticle}
\usepackage{amsmath}
\usepackage{amssymb}
\usepackage{graphicx}

\usepackage{braket}

\usepackage{txfonts}

\usepackage[colorlinks=true,linkcolor=blue,citecolor=blue,urlcolor=blue]{hyperref}

\begin{document}

\title[Entanglement and squeezing in continuous-variable systems]{Entanglement and squeezing\\in continuous-variable systems}
\author{Manuel Gessner}
\email{manuel.gessner@ino.it}
\affiliation{QSTAR, INO-CNR and LENS, Largo Enrico Fermi 2, I-50125 Firenze, Italy}
\author{Luca Pezz\`e}
\affiliation{QSTAR, INO-CNR and LENS, Largo Enrico Fermi 2, I-50125 Firenze, Italy}
\author{Augusto Smerzi}
\affiliation{QSTAR, INO-CNR and LENS, Largo Enrico Fermi 2, I-50125 Firenze, Italy}
\date{\today}

\begin{abstract}
We introduce a multi-mode squeezing coefficient to characterize entanglement in $N$-partite continuous-variable systems. The coefficient relates to the squeezing of collective observables in the $2N$-dimensional phase space and can be readily extracted from the covariance matrix. Simple extensions further permit to reveal entanglement within specific partitions of a multipartite system. Applications with nonlinear observables allow for the detection of non-Gaussian entanglement.
\end{abstract}

\maketitle

\section{Introduction}

Entangled quantum states play a central role in several applications of quantum information theory \cite{NielsenChuang,LucaRMP,Horodecki2009}. Their detection and microscopic characterization, however, poses a highly challenging problem, both theoretically and experimentally \cite{Mintert2005,Horodecki2009,Guhne2009}. 

A rather convenient approach to detect and quantify multipartite entangled states is available for spin systems by the concept of spin squeezing \cite{PhysRevA.46.R6797,PhysRevA.47.5138,Guhne2009,Sorensen2001,PhysRevLett.86.4431,TothPRA2009,SpinSqueezing}. 
Entanglement criteria based on spin-squeezing coefficients are determined by suitable expectation values and variances of collective spin operators. 
This renders these criteria experimentally accessible in atomic systems \cite{LucaRMP}, without the need for local measurements of the individual subsystems. Spin-squeezing based entanglement criteria have been initially introduced for spin-1/2 particles \cite{Sorensen2001, TothPRL2007, TothPRA2009} and 
more recently extended to particles of arbitrary spin \cite{VitaglianoPRL2011} and systems of fluctuating number of particles \cite{PhysRevA.86.012337}.
Furthermore, in Ref.~\cite{Schmied441}, spin squeezing has been related to Bell correlations.
However, so far, entanglement criteria based on spin squeezing have been developed only for discrete-variable systems. A direct extension for continuous-variable systems is not available since the restriction to collective observables prohibits applications in unbounded Hilbert spaces. 

In this article, we construct a bosonic multi-mode squeezing coefficient via a combination of locally and collectively measured variances. 
The additional information provided by local measurements allows us to introduce a generalized squeezing coefficient for arbitrary-dimensional 
systems \cite{ResolutionEnhanced}.
This coefficient is able to detect continuous-variable entanglement and is readily determined on the basis of the covariance matrix. A suitable generalization is further able to reveal entanglement in a specific partition of the multipartite system. As illustrated by examples, our coefficient can be interpreted in terms of squeezed observables in the collective $2N$-dimensional phase space of the $N$-partite system.
We point out that, in continuous-variable systems, local access to the individual modes is routinely provided, 
most prominent examples being multi-mode photonic states \cite{10000modes,PhysRevLett.112.120505,Roslund2014} 
and atomic systems with homodyne measurements \cite{GrossNATURE2011,HamleyNATPHYS2012, Peise2015}.

Our results provide an experimentally feasible alternative to standard entanglement detection techniques based on uncertainty relations or the partial transposition criterion, which are particularly well-suited for Gaussian states \cite{PhysRevLett.84.2726,PhysRevLett.84.2722,PhysRevA.67.022320,RevModPhys.77.513,Adesso07,Weedbrook}. We show how our approach, as well as other variance-based strategies can be interpreted as different modifications of Heisenberg's uncertainty relation, which are only satisfied by separable states. The squeezing coefficient presented here is furthermore directly related to the Fisher information \cite{PhysRevLett.102.100401,PezzePNAS2016,Gessner2016}. In situations where the Fisher information can be extracted experimentally, this relation can be exploited to devise a sharper entanglement criterion \cite{Gessner2016,ResolutionEnhanced}, which is expected to be relevant for the detection of non-Gaussian states.

\section{Entanglement detection with local variances}\label{sec:enhanced}
A unified approach to entanglement detection in arbitrary-dimensional multipartite systems was recently proposed in \cite{Gessner2016}. 
This approach is based on a combination of the Fisher information with suitable local variances. In the following we will briefly review the general ideas. 
Later in this article, this technique will be applied in the context of continuous-variable systems to derive a bosonic squeezing coefficient with 
the ability to detect mode entanglement. 

In an $N$-partite Hilbert space $\mathcal{H}=\mathcal{H}_1\otimes\cdots\otimes\mathcal{H}_N$, any separable quantum state,
\begin{align}\label{eq:fullysep}
\hat{\rho}_{\mathrm{sep}}=\sum_{\gamma}p_{\gamma}\hat{\rho}^{(\gamma)}_1\otimes\cdots\otimes\hat{\rho}^{(\gamma)}_N,
\end{align}
characterized by probabilities $p_{\gamma}$ and local quantum states $\hat{\rho}^{(\gamma)}_i$, must satisfy \cite{Gessner2016}
\begin{align}\label{eq:sepcriterion}
F_Q[\hat{\rho}_{\mathrm{sep}},\hat{A}]&\leq  4\mathrm{Var}(\hat{A})_{\Pi(\hat{\rho}_{\mathrm{sep}})},
\end{align}
where $\hat{A}=\sum_{i=1}^N\hat{A}_i$ is a sum of arbitrary local operators $\hat{A}_i$ that act on the $\mathcal{H}_i$, respectively. We have further introduced the quantum mechanical variance $\mathrm{Var}(\hat{A})_{\hat{\rho}}=\langle\hat{A}^2\rangle_{\hat{\rho}} -\langle\hat{A}\rangle_{\hat{\rho}}^2$ and expectation value $\langle \hat{A}\rangle_{\hat{\rho}}=\mathrm{Tr}[\hat{A}\hat{\rho}]$. The uncorrelated product state $\Pi(\hat{\rho})=\hat{\rho}_1\otimes\cdots\otimes\hat{\rho}_N$ is constructed from the reduced density operators $\hat{\rho}_i$ of $\hat{\rho}$, i.e., it describes the state $\hat{\rho}$ after the removal of all correlations between the subsystems $\mathcal{H}_i$, for $i=1,\dots,N$.

On the left-hand side of Eq.~(\ref{eq:sepcriterion}) appears the quantum Fisher information $F_Q[\hat{\rho},\sum_{i=1}^N \hat{A}_i]$, which expresses how sensitively the state $\hat{\rho}(\theta)=e^{-i\sum_{j=1}^N \hat{A}_j\theta} \hat{\rho} e^{i\sum_{j=1}^N \hat{A}_j\theta}$ changes upon small variations of $\theta$ \cite{PhysRevLett.72.3439,paris2009,Giovannetti2011,Varenna}. This quantity is compared to the sum of local variances of the $\hat{A}_i$ that generate the unitary transformation $e^{-i\sum_{j=1}^N \hat{A}_j\theta}$: The term on the right-hand side can be expressed as $\mathrm{Var}(\sum_{i=1}^N\hat{A}_i)_{\Pi(\hat{\rho})}=\sum_{i=1}^N\mathrm{Var}\big(\hat{A}_i\big)_{\hat{\rho}}$. The combination of the Fisher information with local variances provides two decisive advantages. First, it renders the separability criterion~(\ref{eq:sepcriterion}) stricter than those that are based on a state-dependent upper bound for the local variances \cite{Gessner2016,ResolutionEnhanced}. Indeed, the condition~(\ref{eq:sepcriterion}) is necessary and sufficient for separability of pure states: for every 
pure entangled state it is possible to find a set of local opertors $\hat{A}_i$ such that Eq.~(\ref{eq:sepcriterion}) is violated \cite{Gessner2016}. Second, the local variances lead to a finite bound for the Fisher information, even in the presence of unbounded operators. This is of particular importance for applications in the context of continuous-variable systems.

The Fisher information has been extracted in certain (discrete-variable) experiments with a state-independent method \cite{Strobel424,Bohnet1297}. In the present article we focus mostly on simpler entanglement criteria that are available directly from the covariance matrix of the quantum state. To this end, we make use of the upper bound \cite{PhysRevLett.102.100401,Varenna}
\begin{align}\label{eq:FBound}
F_Q[\hat{\rho},\hat{A}]\geq \frac{|\langle [\hat{A},\hat{B}]\rangle_{\hat{\rho}}|^2}{\mathrm{Var}(\hat{B})_{\hat{\rho}}},
\end{align}
which holds for arbitrary states $\hat{\rho}$ and an arbitrary pair of operators $\hat{A}$, $\hat{B}$. In combination with Eq.~(\ref{eq:sepcriterion}), we obtain the variance-based separability criterion \cite{Gessner2016,ResolutionEnhanced}
\begin{align}\label{eq:variancecriterion}
\mathrm{Var}(\hat{A})_{\Pi(\hat{\rho}_{\mathrm{sep}})}\mathrm{Var}(\hat{B})_{\hat{\rho}_{\mathrm{sep}}}\geq \frac{|\langle [\hat{A},\hat{B}]\rangle_{\hat{\rho}_{\mathrm{sep}}}|^2}{4},
\end{align}
where $\hat{B}$ is an arbitrary operator and $\hat{A}=\sum_{i=1}^N\hat{A}_i$, as before. Based on the separability condition~(\ref{eq:variancecriterion}) a family of entanglement witnesses may be formulated in terms of generalized squeezing coefficients \cite{ResolutionEnhanced}
\begin{align}\label{eq:genSC}
\xi^2_{\hat{A},\hat{B}}(\hat{\rho})=\frac{4\mathrm{Var}(\hat{A})_{\Pi(\hat{\rho})}\mathrm{Var}(\hat{B})_{\hat{\rho}}}{|\langle [\hat{A},\hat{B}]\rangle_{\hat{\rho}}|^2}.
\end{align}
For separable states, the attainable values of $\xi^2_{\hat{A},\hat{B}}$ are bounded by
\begin{align}\label{eq:sccrit}
\xi^{2}_{\hat{A},\hat{B}}(\hat{\rho}_{\mathrm{sep}})\geq 1,
\end{align}
as a consequence of Eq.~(\ref{eq:variancecriterion}). Since this bound holds for arbitrary choices of $\hat{A}=\sum_{i=1}^N\hat{A}_i$ and $\hat{B}$, these operators can be optimized to obtain the most efficient entanglement witness. In the case of spin systems, Eq.~(\ref{eq:genSC}) indeed reproduces the spin-squeezing coefficient, and several sharpened generalizations thereof as special cases \cite{ResolutionEnhanced}. In Sec.~\ref{sec:CV}, we will derive such a coefficient for bosonic continuous-variable systems.

We further remark that the coefficient~(\ref{eq:genSC}) can be made more sensitive to entanglement if knowledge of the Fisher information is available \cite{PhysRevLett.102.100401,Gessner2016,ResolutionEnhanced}. The variance-assisted Fisher density \cite{ResolutionEnhanced}
\begin{align}\label{eq:genFD}
f_{\hat{A}}(\hat{\rho})=\frac{F_Q[\hat{\rho},\hat{A}]}{4\mathrm{Var}(\hat{A})_{\Pi(\hat{\rho})}},
\end{align}
satisfies $f_{\hat{A}}(\hat{\rho}_{\mathrm{sep}})\leq 1$ for separable states $\hat{\rho}_{\mathrm{sep}}$, due to Eq.~(\ref{eq:sepcriterion}). This criterion is stronger than Eq.~(\ref{eq:sccrit}) since $f_{\hat{A}}(\hat{\rho})\geq \xi^{-2}_{\hat{A},\hat{B}}(\hat{\rho})$ holds as a consequence of Eq.~(\ref{eq:FBound}). We will discuss in Sec.~\ref{sec:multi} how both coefficients~(\ref{eq:genSC}) and~(\ref{eq:genFD}) can be adjusted to reveal entanglement in a specific partition, rather than just anywhere in the system.

\section{Multi-mode continuous variable systems}\label{sec:CV}
We now apply the above entanglement criteria to a bosonic continuous-variable system of $N$ modes. We will focus particularly on the squeezing coefficient~(\ref{eq:genSC}). The associated entanglement criterion~(\ref{eq:variancecriterion}) manifests as a modification of Heisenberg's uncertainty relation for separable states, allowing us to compare our method to other existing techniques. The optimal choice of the operators $\hat{A}$ and $\hat{B}$ in Eq.~(\ref{eq:genSC}) can be found in terms of $2N$-dimensional phase space vectors. These lead to a geometrical interpretation and associate the coefficient to multi-mode squeezing in phase space.

\subsection{Comparison to related entanglement criteria and the uncertainty principle}
We begin by comparing the criterion~(\ref{eq:variancecriterion}) to current state-of-the-art criteria for continuous-variable systems. 
Most of the existing entanglement criteria for continuous variables are formulated in terms of second moments, i.e., functions of the covariance matrix \cite{PhysRevA.67.052315,PhysRevA.90.052321,PhysRevA.90.062337,PhysRevA.92.042328,PhysRevLett.84.2722,PhysRevLett.84.2726,PhysRevLett.114.050501}. Standard methods for entanglement detection are based on Gaussian tests of the positive partial transposition (PPT) criterion \cite{PhysRevLett.84.2726}, which are applicable to bipartite systems and yield a necessary and sufficient condition for separability of two-mode Gaussian states \cite{PhysRevLett.84.2722,PhysRevLett.84.2726}. If both subsystems consist of more than one mode, this condition is no longer sufficient as bound entanglement can arise \cite{WernerWolf}. The PPT criterion is further intimately related to separability conditions \cite{PhysRevA.67.022320} based on Heisenberg-Robertson uncertainty relations \cite{Heisenberg1927,PhysRev.34.163}, which can be sharpened with an entropic formulation of the uncertainty principle \cite{PhysRevLett.103.160505}; see also \cite{Yichen}.

The most general formulation of these criteria for bipartite systems was provided in Ref.~\cite{PhysRevA.67.022320}. We point out that the proof presented there can be straight-forwardly generalized to a multipartite scenario (see Appendix~\ref{app:giovannettimulti}), yielding the separability condition
\begin{align}\label{eq:MultiGiovannetti}
&\quad\mathrm{Var}(\hat{A}(\boldsymbol{\alpha}))_{\hat{\rho}_{\mathrm{sep}}}\mathrm{Var}(\hat{B}(\boldsymbol{\beta}))_{\hat{\rho}_{\mathrm{sep}}}\notag\\&\geq\frac{\left(\sum_{i=1}^N\left|\alpha_i\beta_i\langle[\hat{A}_i,\hat{B}_i]\rangle_{\hat{\rho}_{\mathrm{sep}}}\right|\right)^2}{4},
\end{align}
where $\hat{A}(\boldsymbol{\alpha})=\sum_{i=1}^N\alpha_i\hat{A}_i$ and $\hat{B}(\boldsymbol{\beta})=\sum_{i=1}^N\beta_i\hat{B}_i$ are sums of arbitrary operators $\hat{A}_i$ and $\hat{B}_i$ that act on the local Hilbert spaces $\mathcal{H}_i$, with real coefficients $\boldsymbol{\alpha}=(\alpha_1,\dots,\alpha_N)$ and $\boldsymbol{\beta}=(\beta_1,\dots,\beta_N)$. For two-mode states ($N=2$), the criterion~(\ref{eq:MultiGiovannetti}) was shown \cite{PhysRevA.67.022320} to be stronger than criteria that employ the sum (rather than the product) of two variances \cite{PhysRevLett.84.2722}, which in turn were extended to multipartite systems in \cite{PhysRevA.67.052315}; see also \cite{PhysRevA.90.062337,PhysRevLett.117.140504}. When the operators $\hat{A}_i$ and $\hat{B}_i$ are limited to quadratures, the family of criteria~(\ref{eq:MultiGiovannetti}), for $N=2$, is equivalent to the PPT criterion \cite{PhysRevA.67.022320}.

The criterion~(\ref{eq:MultiGiovannetti}) can now be compared to the entanglement criteria that follow from the approach presented in Sec.~\ref{sec:enhanced} in the case of continuous-variable systems. Following Eq.~(\ref{eq:variancecriterion}) we find that any separable continuous-variable state $\hat{\rho}_{\mathrm{sep}}$ must obey the bound
\begin{align}\label{eq:CVVariance}
&\quad\mathrm{Var}(\hat{A}(\boldsymbol{\alpha}))_{\Pi(\hat{\rho}_{\mathrm{sep}})}\mathrm{Var}(\hat{B}(\boldsymbol{\beta}))_{\hat{\rho}_{\mathrm{sep}}}\notag\\&\geq\frac{\left|\sum_{i=1}^N\alpha_i\beta_i\langle[\hat{A}_i,\hat{B}_i]\rangle_{\hat{\rho}_{\mathrm{sep}}}\right|^2}{4}.
\end{align}
In contrast to Eq.~(\ref{eq:MultiGiovannetti}), the uncertainty principle is not used to derive Eq.~(\ref{eq:CVVariance}). We further remark that, while in Eq.~(\ref{eq:MultiGiovannetti}), the operator $\hat{B}(\boldsymbol{\beta})$ needs to be a sum of local operators, the criterion~(\ref{eq:CVVariance}) can be sharpened by using more general operators instead of $\hat{B}(\boldsymbol{\beta})$.

The two necessary conditions for separability~(\ref{eq:MultiGiovannetti}) and (\ref{eq:CVVariance}) should be compared to the general bound
\begin{align}\label{eq:Heisenberg}
&\quad\mathrm{Var}(\hat{A}(\boldsymbol{\alpha}))_{\hat{\rho}}\mathrm{Var}(\hat{B}(\boldsymbol{\beta}))_{\hat{\rho}}\notag\\&\geq\frac{\left|\sum_{i=1}^N\alpha_i\beta_i\langle[\hat{A}_i,\hat{B}_i]\rangle_{\hat{\rho}}\right|^2}{4},
\end{align}
provided by the Heisenberg-Robertson uncertainty relation for \textit{arbitrary} states \cite{Heisenberg1927,PhysRev.34.163}. We observe that both methods can be interpreted as restrictions on the uncertainty relation, which lead to different conditions that are only satisfied by separable states. On the one hand, in Eq.~(\ref{eq:MultiGiovannetti}) the modification of Eq.~(\ref{eq:Heisenberg}) is given by a tighter upper bound for separable states, $(\sum_{i=1}^N|\alpha_i\beta_i\langle[\hat{A}_i,\hat{B}_i]\rangle_{\hat{\rho}}|)^2\geq|\sum_{i=1}^N\alpha_i\beta_i\langle[\hat{A}_i,\hat{B}_i]\rangle_{\hat{\rho}}|^2$. On the other hand, in Eq.~(\ref{eq:CVVariance}), the variance $\mathrm{Var}(\hat{A}(\boldsymbol{\alpha}))_{\Pi(\hat{\rho}_{\mathrm{sep}})}$ is obtained for the product state $\Pi(\hat{\rho}_{\mathrm{sep}})$, and thus differs from the uncertainty relation~(\ref{eq:Heisenberg}), despite the fact that the right-hand side of the two inequalities coincide.

In the following we focus on the separability condition~(\ref{eq:CVVariance}). We reformulate Eq.~(\ref{eq:CVVariance}) in terms of multidimensional quadratures (Sec.~\ref{sec:multidimquad}) and covariance matrices, which leads to the definition of a multi-mode squeezing coefficient (Sec.~\ref{sec:mmsqueeze}).

\subsection{Formulation in terms of multidimensional quadratures}\label{sec:multidimquad}
Let us briefly illustrate how the criteria~(\ref{eq:CVVariance}) can be used with specific choices of accessible operators. In the discrete-variable case, the most general non-trivial local operator can be parametrized in terms of a finite number of generating operators, allowing one to systematically optimize the spin-squeezing coefficient as a function of these parameters \cite{TothPRA2009,SpinSqueezing,ResolutionEnhanced}. In infinite-dimensional Hilbert spaces such a parametrization would involve an infinite number of parameters. The entanglement criteria that can be constructed using Eqs.~(\ref{eq:sepcriterion}) and~(\ref{eq:CVVariance}) therefore depend on a predefined set of accessible operators \cite{Gessner2016}. 

A common choice for such a set are the local position $\hat{x}_1,\dots,\hat{x}_N$ and momentum operators $\hat{p}_1,\dots,\hat{p}_N$, or more general quadratures of the type $\hat{q}_j(\theta_j)=\frac{1}{\sqrt{2}}\left(e^{i\theta_j}\hat{a}_j^{\dagger}+e^{-i\theta_j}\hat{a}_j\right)$. Here, $\hat{a}_i$ is the annihilation operator of the mode $i$ and the quadrature $\hat{q}_j({\theta_j})$ includes the special cases $\hat{q}_j(0)=\hat{x}_j$ and $\hat{q}_j(\pi/2)=\hat{p}_j$. 

Let us consider
\begin{align}
\hat{M}(\mathbf{v})&=\sum_{i=1}^N\left(n_i\hat{x}_i+m_i\hat{p}_i\right)=\sum_{i=1}^N\frac{1}{\sqrt{2}}\left(v_i\hat{a}^{\dagger}_i+v_i^*\hat{a}_i\right),
\end{align}
where $\mathbf{v}=(v_1,\dots,v_N)$, $v_j=n_j+im_j$, and the coefficients $n_j$ and $m_j$ are chosen real. We obtain the following separability criterion from~(\ref{eq:CVVariance}):
\begin{align}\label{eq:quadraturescriterion}
\mathrm{Var}(\hat{M}(\mathbf{v}))_{\Pi(\hat{\rho}_{\mathrm{sep}})}\mathrm{Var}(\hat{M}(\mathbf{w}))_{\hat{\rho}_{\mathrm{sep}}}&\geq\frac{\left|\sum_{i=1}^N\mathrm{Im}(v^{*}_iw_i)\right|^2}{4}
\end{align}
where $\mathrm{Im}$ denotes the imaginary part. 

As a special case we may constrain both $\mathbf{v}$ and $\mathbf{w}$ to the unit circle: $v_j=e^{i\theta_j}$ and $w_j=e^{i\phi_j}$. We then obtain the multi-mode quadrature operators, e.g., 
\begin{align}
\hat{Q}_{\boldsymbol{\theta}}=\sum_{j=1}^N\hat{q}_j(\theta_j)=\frac{1}{\sqrt{2}}\sum_{j=1}^N(e^{i\theta_j}\hat{a}^{\dagger}_j+e^{-i\theta_j}\hat{a}_j),
\end{align}
and with Eq.~(\ref{eq:quadraturescriterion}), we find the separability bound
\begin{align}
\mathrm{Var}(\hat{Q}_{\boldsymbol{\theta}})_{\Pi(\hat{\rho}_{\mathrm{sep}})}\mathrm{Var}(\hat{Q}_{\boldsymbol{\phi}})_{\hat{\rho}_{\mathrm{sep}}}&\geq\frac{\left|\sum_{i=1}^N\sin(\theta_i-\phi_i)\right|^2}{4}.
\end{align}
Alternatively, $\mathbf{v}$ and $\mathbf{w}$ may be chosen purely imaginary and purely real, respectively. Introducing the notation
\begin{align}\label{eq:multidimXP}
\hat{X}_{\mathbf{n}}&=\sum_{i=1}^Nn_i\hat{x}_i, \qquad \hat{P}_{\mathbf{m}}=\sum_{i=1}^Nm_i\hat{p}_i,
\end{align}
we obtain the separability condition
\begin{align}\label{eq:xpcondition}
\mathrm{Var}(\hat{X}_{\mathbf{n}})_{\Pi(\hat{\rho}_{\mathrm{sep}})}\mathrm{Var}(\hat{P}_{\mathbf{m}})_{\hat{\rho}_{\mathrm{sep}}}
\geq\frac{\left|\mathbf{n}\cdot\mathbf{m}\right|^2}{4},
\end{align}
where $\mathbf{n}\cdot\mathbf{m}=\sum_{i=1}^Nn_im_i$, and the roles of $\hat{X}_{\mathbf{n}}$ and $\hat{P}_{\mathbf{m}}$ may be exchanged.

So far, we restricted to operators of first order in $\hat{a}_i$ and $\hat{a}^{\dagger}_i$. An example of a second-order operator is given by the number operator $\hat{N}=\sum_{i=1}^N\hat{n}_i$. The commutation relations
\begin{align}
[\hat{X}_{\mathbf{m}},\hat{N}]=i\hat{P}_{\mathbf{m}},\qquad [\hat{P}_{\mathbf{m}},\hat{N}]=i\hat{X}_{\mathbf{m}},
\end{align}
however, render these combinations of operators less suitable for entanglement detection. The reason is that the expectation values of $\hat{X}_{\mathbf{m}}$ and $\hat{P}_{\mathbf{m}}$, which play the role of the upper separability bound in Eq.~(\ref{eq:CVVariance}), can be set to zero with local operations. A family of second-order quadrature operators will be discussed in Sec.~\ref{sec:nlsq}.

\subsection{Multi-mode squeezing coefficient as entanglement witness}\label{sec:mmsqueeze}
In general, the criteria of the type~(\ref{eq:CVVariance}) are conveniently expressed using the covariance formalism of continuous variables \cite{Paris2005,RevModPhys.77.513,Adesso07,Weedbrook}. Introducing the $2N$-dimensional vector $\hat{\mathbf{r}}=(\hat{r}_1,\dots,\hat{r}_{2N})=(\hat{x}_1,\hat{p}_1,\dots,\hat{x}_N,\hat{p}_N)$, we define the covariance matrix of the state $\hat{\rho}$ element-wise as 
\begin{align}
(\boldsymbol{\gamma}_{\hat{\rho}})_{\alpha\beta}=\mathrm{Cov}(\hat{r}_{\alpha},\hat{r}_{\beta})_{\hat{\rho}}.
\end{align}
Arbitrary local operators
\begin{align}\label{eq:Mofc}
\hat{M}(\mathbf{g})=\sum_{i=1}^{2N}g_i\hat{r}_i,
\end{align}
with the real vector $\mathbf{g}=(g_1,\dots,g_{2N})$ now generate the commutation relations
\begin{align}\label{eq:MMcommut}
[\hat{M}(\mathbf{h}),\hat{M}(\mathbf{g})]=\sum_{i,j=1}^{2N}h_ig_j[\hat{r}_i,\hat{r}_j]=i\mathbf{h}^T\boldsymbol{\Omega}\mathbf{g},
\end{align}
where $\boldsymbol{\Omega}=\bigoplus_{i=1}^N\boldsymbol{\omega}$ is the symplectic form with $\omega=\left(\begin{smallmatrix} 0 & 1\\-1 & 0 \end{smallmatrix}\right)$. Furthermore, the variances of such operators are given in terms of the covariance matrix as:
\begin{align}\label{eq:Mofgvar}
\mathrm{Var}(\hat{M}(\mathbf{g}))_{\hat{\rho}}=\sum_{i,j=1}^{2N}g_ig_j\mathrm{Cov}(\hat{r}_i,\hat{r}_j)_{\hat{\rho}}=\mathbf{g}^T\boldsymbol{\gamma}_{\hat{\rho}}\mathbf{g}.
\end{align}
We can thus rewrite the separability criterion Eq.~(\ref{eq:CVVariance}) in the equivalent form
\begin{align}\label{eq:covariancecriterion}
(\mathbf{h}^T\boldsymbol{\gamma}_{\Pi(\hat{\rho}_{\mathrm{sep}})}\mathbf{h})(\mathbf{g}^T\boldsymbol{\gamma}_{\hat{\rho}_{\mathrm{sep}}}\mathbf{g})\geq \frac{\left(\mathbf{h}^T\boldsymbol{\Omega}\mathbf{g}\right)^2}{4},
\end{align}
where the correlation-free covariance matrix $\boldsymbol{\gamma}_{\Pi(\hat{\rho}_{\mathrm{sep}})}$ is obtained from $\boldsymbol{\gamma}_{\hat{\rho}_{\mathrm{sep}}}$ by setting all elements to zero except for the local $2\times 2$ blocks on the diagonal. Again, we may compare Eq.~(\ref{eq:covariancecriterion}) to the Heisenberg-Robertson uncertainty relation~(\ref{eq:Heisenberg}), which here reads
\begin{align}\label{eq:HeisenbergCOV}
(\mathbf{h}^T\boldsymbol{\gamma}_{\hat{\rho}}\mathbf{h})(\mathbf{g}^T\boldsymbol{\gamma}_{\hat{\rho}}\mathbf{g})\geq \frac{\left(\mathbf{h}^T\boldsymbol{\Omega}\mathbf{g}\right)^2}{4},
\end{align}
and holds for arbitrary $\hat{\rho}$.

The number of degrees of freedom in Eq.~(\ref{eq:covariancecriterion}) can be halved by considering only pairs of operators that maximize the right-hand side, which is given by the scalar product of the two vectors $\mathbf{h}$ and $\boldsymbol{\Omega}\mathbf{g}$. For two vectors of fixed length, the scalar product reaches its maximum value when they are parallel or anti-parallel, hence, if $\mathbf{h}=\pm\boldsymbol{\Omega}\mathbf{g}$ (recall also that $\boldsymbol{\Omega}\boldsymbol{\Omega}=-\mathbb{I}_{2N}$). Generally, we call the directions $\pm\boldsymbol{\Omega}\mathbf{g}$ \textit{maximally non-commuting} with respect to $\mathbf{g}$ since pairs of operators $\hat{M}(\mathbf{g})$ and $\hat{M}(\pm\boldsymbol{\Omega}\mathbf{g})$ maximize the absolute value of Eq.~(\ref{eq:MMcommut}). This yields the following necessary condition for separability
\begin{align}\label{eq:simplecovcrit}
\xi^{2}(\hat{\rho}_{\mathrm{sep}})\geq 1,
\end{align}
where, based on the general formulation~(\ref{eq:genSC}), we introduced the \textit{bosonic multi-mode squeezing coefficient}
\begin{align}\label{eq:chi2}
\xi^2(\hat{\rho}):=\min_{\mathbf{g}}\xi^2_{\mathbf{g}}(\hat{\rho}),
\end{align}
with
\begin{align}\label{eq:chig}
\xi^2_{\mathbf{g}}(\hat{\rho}):=\frac{4(\mathbf{g}^T\boldsymbol{\Omega}^T\boldsymbol{\gamma}_{\Pi(\hat{\rho})}\boldsymbol{\Omega}\mathbf{g})(\mathbf{g}^T\boldsymbol{\gamma}_{\hat{\rho}}\mathbf{g})}{(\mathbf{g}^T\mathbf{g})^2}.
\end{align}
Notice that $\xi^2_{\mathbf{g}}$ does not depend on the normalization of $\mathbf{g}$.

The coefficient $\xi^2_{\mathbf{g}}(\hat{\rho})$ can be immediately obtained from any given covariance matrix. The optimization involved in Eq.~(\ref{eq:chi2}) contains $2N$ free parameters, whereas the normalization of $\mathbf{g}$ reduces this number by one. In the two examples discussed below, we notice that a reasonable point of departure for finding the $\mathbf{g}$ that minimizes Eq.~(\ref{eq:chi2}), is to choose $\mathbf{g}$ as an eigenvector of $\boldsymbol{\gamma}_{\hat{\rho}}$ with minimal eigenvalue, i.e., to focus first on the minimization of the last factor in Eq.~(\ref{eq:chig}).

The coefficients $\xi^2_{\mathbf{g}}(\hat{\rho})$ allow for an interpretation that creates a notion of bosonic multi-mode squeezing and relates it to entanglement. To this end, one interprets the vectors $\mathbf{g}$ as `directions' in the $2N$-dimensional phase space. A small value of $\mathbf{g}^T\boldsymbol{\gamma}_{\hat{\rho}}\mathbf{g}$ can be interpreted as a squeezed variance along the direction $\mathbf{g}$ (here considered to be normalized to one), as by Eq.~(\ref{eq:Mofgvar}) it reflects a small variance of the operator $\hat{M}(\mathbf{g})$, in close analogy to the case of spins \cite{ResolutionEnhanced}. In order to satisfy Eq.~(\ref{eq:HeisenbergCOV}), the squeezing along $\mathbf{g}$ entails anti-squeezing along the maximally non-commuting direction $\boldsymbol{\Omega}\mathbf{g}$. If the anti-squeezing along $\boldsymbol{\Omega}\mathbf{g}$ can be suppressed by removing mode correlations (as is done by the operation $\Pi$), it is possible to achieve small values of $\xi^2$, and, in the presence of mode entanglement, to violate the bound~(\ref{eq:simplecovcrit}).

\subsection{Examples}
We now discuss the examples of $N$-mode squeezed states for $N=2,3$ and show that they violate the separability criterion~(\ref{eq:simplecovcrit}) for any finite squeezing. Moreover, the entanglement of a non-Gaussian squeezed state is detected with the aid of nonlinear observables.

\subsubsection{Two-mode squeezed states}\label{sec:2ms}
We first consider two-mode squeezed vacuum states 
\begin{align}
|\Psi^{(2)}_r\rangle=\hat{S}_{12}[r]|0,0\rangle,
\end{align}
generated by the operation $\hat{S}_{12}[\xi]=e^{\xi \hat{a}_1^{\dagger}\hat{a}_2^{\dagger}-\xi^*\hat{a}_1\hat{a}_2}$ \cite{Paris2005,RevModPhys.77.513,Adesso07} and $|0\rangle$ is the vacuum. One confirms easily that the states $|\Psi^{(2)}_r\rangle$ violate Eq.~(\ref{eq:simplecovcrit}), which reveals their inseparability. To see this, we notice first that these states are Gaussian, and therefore fully characterized by their covariance matrix
\begin{align} \label{eq:CMmatrix2}
\boldsymbol{\gamma}_{|\Psi^{(2)}_r\rangle}=
\frac{1}{2}\begin{pmatrix} 
R^{(2)} & 0 & S^{(2)} & 0 \\
0 & R^{(2)} & 0 & -S^{(2)} \\ 
S^{(2)} & 0 & R^{(2)} & 0 \\ 
0 & -S^{(2)} & 0 & R^{(2)} 
\end{pmatrix},
\end{align}
where $R^{(2)} = \cosh(2r)$ and $S^{(2)} = \sinh(2r)$. For $r>0$, it describes position correlations and momentum anti-correlations: $\mathrm{Var}(\hat{x}_1-\hat{x}_2)_{|\Psi^{(2)}_r\rangle}=\mathrm{Var}(\hat{p}_1+\hat{p}_2)_{|\Psi^{(2)}_r\rangle}=e^{-2r}$. We take $\mathbf{g}_{xp}=(c_x,c_p,-c_x,c_p)$ with arbitrary numbers $c_x$ and $c_p$, 
which is the eigenvector of the matrix (\ref{eq:CMmatrix2}) with minimum eigenvalue $( R^{(2)} - S^{(2)})/2$.
This choice corresponds to the operator $\hat{M}(\mathbf{g}) = c_x(\hat{x}_1-\hat{x}_2) + c_p(\hat{p}_1+\hat{p}_2)$ and leads to
$\mathbf{g}^T \boldsymbol{\gamma}_{|\Psi^{(2)}_r\rangle} \mathbf{g} 
=(c_x^2 + c_p^2)e^{-2 r}$. We can obtain the correlation-free covariance matrix from Eq.~(\ref{eq:CMmatrix2}) by setting all elements except the diagonal $2\times 2$-blocks to zero, yielding
\begin{align} \label{eq:CMmatrix2lcl}
\boldsymbol{\gamma}_{\Pi(|\Psi^{(2)}_r\rangle)}=
\frac{1}{2}\begin{pmatrix} 
R^{(2)} & 0 & 0 & 0 \\
0 & R^{(2)} & 0 & 0 \\ 
0 & 0 & R^{(2)} & 0 \\ 
0 & 0 & 0 & R^{(2)} 
\end{pmatrix}.
\end{align}
Thus, $\mathbf{g}_{xp}^T\boldsymbol{\Omega}^T\boldsymbol{\gamma}_{\Pi(|\Psi^{(2)}_r\rangle)}\boldsymbol{\Omega}\mathbf{g}_{xp}=(c_x^2+c_p^2)R^{(2)}$ and with $\mathbf{g}_{xp}^T\mathbf{g}_{xp}=2(c_x^2+c_p^2)$, we finally find
\begin{align}\label{eq:tmsa}
\xi^{2}_{\mathbf{g}_{xp}}(|\Psi^{(2)}_r\rangle)=\frac{1}{2}(1+e^{-4r}),
\end{align}
which violates Eq.~(\ref{eq:simplecovcrit}) for all $r>0$. 

\begin{figure}[tb]
\centering
\includegraphics[width=.45\textwidth]{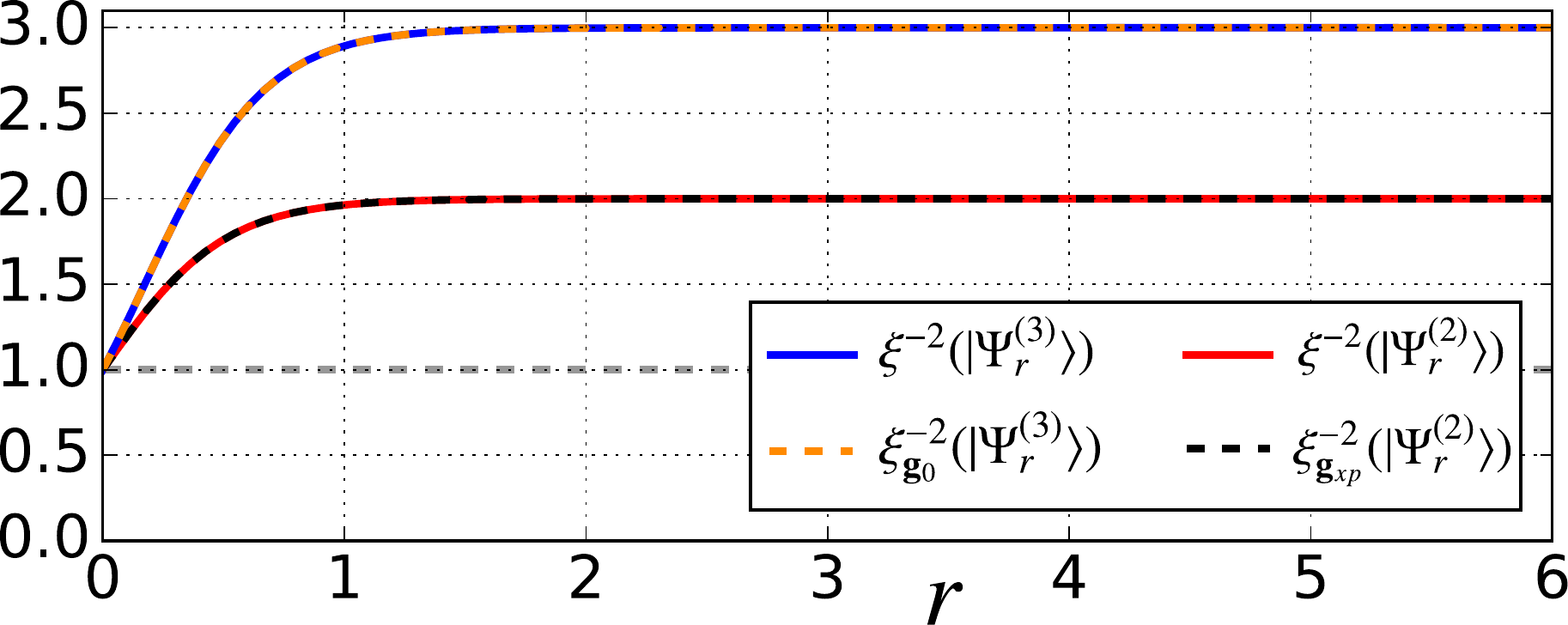}
\caption{Numerically optimized inverse multi mode squeezing coefficients~(\ref{eq:chi2}) and analytical predictions for particular directions in the $2N$-dimensional phase space according to Eqs.~(\ref{eq:tmsa}) and~(\ref{eq:3msa}) for the two- and three-mode squeezed states $|\Psi_r^{(2)}\rangle$ and $|\Psi_r^{(3)}\rangle$, respectively. The coefficients violate the separability condition~(\ref{eq:simplecovcrit}) for all values of the squeezing parameter $r>0$.}
\label{fig:CVMMS}
\end{figure}

The choice of $\mathbf{g}_{xp}$ is indeed confirmed as optimal by observing in Fig.~\ref{fig:CVMMS} that $\xi^{-2}(|\Psi^{(2)}_r\rangle)=\xi^{-2}_{\mathbf{g}_{xp}}(|\Psi^{(2)}_r\rangle)$, where $\xi^{-2}(|\Psi^{(2)}_r\rangle)$ was obtained by minimizing the quantity~(\ref{eq:chi2}) numerically. 

\subsubsection{Three-mode squeezed states}\label{sec:3ms}
A three-mode squeezed state can be generated by single-mode vacuum squeezing of all three modes followed by two consecutive two-mode mixing operations \cite{PhysRevLett.91.080404}. Specifically, we consider the states
\begin{align}
|\Psi^{(3)}_r\rangle&=\hat{B}_{23}\left[\frac{\pi}{4}\right]\hat{B}_{12}\left[\arccos\left(\frac{1}{\sqrt{3}}\right)\right]\notag\\&\quad\times\hat{S}_3[r]\hat{S}_2[r]\hat{S}_1[-r]|0,0,0\rangle,
\end{align}
where $\hat{S}_i[\xi]=e^{\frac{1}{2}(\xi \hat{a}_i^{\dagger 2}-\xi^*\hat{a}_i^{2})}$ is the single-mode squeezing operator of mode $i$ and $\hat{B}_{ij}[\theta]=e^{(\hat{a}_i\hat{a}_j^{\dagger}-\hat{a}_i^{\dagger}\hat{a}_j)\theta}$ mixes the modes $i$ and $j$ with angle $\theta\in\mathbb{R}$. These states are also Gaussian, with covariance matrix \cite{Paris2005}
\begin{align} \label{eq:3mscov}
&\qquad\boldsymbol{\gamma}_{|\Psi^{(3)}_r\rangle}\notag\\&=\frac{1}{2}\begin{pmatrix} 
R_+^{(3)} & 0 & S^{(3)} & 0 & S^{(3)} & 0
\\ 0 & R_-^{(3)} & 0 & -S^{(3)} & 0 & -S^{(3)}
\\ S^{(3)} & 0 & R_+^{(3)} & 0 & S^{(3)} & 0
\\ 0 & -S^{(3)} & 0 & R_-^{(3)} & 0 & -S^{(3)}
\\ S^{(3)} & 0 & S^{(3)} & 0 & R^{(3)}_+ & 0
\\ 0 & -S^{(3)} & 0 & -S^{(3)} & 0 & R_-^{(3)}
\end{pmatrix},
\end{align}
where $R^{(3)}_{\pm}=\cosh(2r)\pm\frac{1}{3}\sinh(2r)$ and $S^{(3)}=-\frac{2}{3}\sinh(2r)$. The eigenspace of the lowest eigenvalue $(1/2)e^{-2r}$ is spanned by the three vectors $\mathbf{g}_0=(1,0,1,0,1,0)$, $\mathbf{g}_1=(0,-1,0,-1,0,2)$ and $\mathbf{g}_2=(0,1,0,-1,0,0)$. These vectors thus identify $2N$-dimensional directions in phase space along which the states $|\Psi^{(3)}_r\rangle$ are squeezed. Let us first focus on $\mathbf{g}_0$. A maximally non-commuting direction with respect to $\mathbf{g}_0$ is given by $-\boldsymbol{\Omega}\mathbf{g}_0=(0,1,0,1,0,1)$. This identifies a direction of anti-squeezing in phase space since the vector $-\boldsymbol{\Omega}\mathbf{g}_0$ is a (non-normalized) eigenvector of $\boldsymbol{\gamma}_{|\Psi^{(3)}_r\rangle}$ with maximal eigenvalue $(1/2)e^{2r}$. This, of course, is required to satisfy the uncertainty relation~(\ref{eq:HeisenbergCOV}). The particular form of the directions $\mathbf{g}_0$ and $-\boldsymbol{\Omega}\mathbf{g}_0$ allows for an interpretation in terms of $N$-dimensional position and momentum quadratures, as introduced in Eq.~(\ref{eq:multidimXP}). We can identify $N$-dimensional position squeezing along $\mathbf{m}_0=(1,1,1)$ with
\begin{align}\label{eq:3modeX}
\mathrm{Var}(\hat{X}_{\mathbf{m}_0})_{|\Psi^{(3)}_r\rangle}=\mathbf{g}_0^T\boldsymbol{\gamma}_{|\Psi^{(3)}_r\rangle}\mathbf{g}_0=\frac{3}{2}e^{-2r},
\end{align}
and momentum anti-squeezing with
\begin{align}
\mathrm{Var}(\hat{P}_{\mathbf{m}_0})_{|\Psi^{(3)}_r\rangle}=\mathbf{g}_0^T\boldsymbol{\Omega}\boldsymbol{\gamma}_{|\Psi^{(3)}_r\rangle}\boldsymbol{\Omega}\mathbf{g}_0=\frac{3}{2}e^{2r}.
\end{align}

However, for the separability condition~(\ref{eq:covariancecriterion}), we compare to the correlation-free covariance matrix, again obtained by removing the off-diagonal correlation blocks $\boldsymbol{\gamma}_{|\Psi^{(3)}_r\rangle}$, leading to
\begin{align} 
&\qquad\boldsymbol{\gamma}_{\Pi(|\Psi^{(3)}_r\rangle)}\notag\\&=\frac{1}{2}\begin{pmatrix} 
R_+^{(3)} & 0 & 0 & 0 & 0 & 0
\\ 0 & R_-^{(3)} & 0 & 0 & 0 & 0
\\ 0 & 0 & R_+^{(3)} & 0 & 0 & 0
\\ 0 & 0 & 0 & R_-^{(3)} & 0 & 0
\\ 0 & 0 & 0 & 0 & R^{(3)}_+ & 0
\\ 0 & 0 & 0 & 0 & 0 & R_-^{(3)}
\end{pmatrix}.
\end{align}
The removal of the correlations has indeed suppressed the anti-squeezing along $-\boldsymbol{\Omega}\mathbf{g}_0$, or equivalently for $\hat{P}_{\mathbf{m}_0}$: We find 
\begin{align}
\mathrm{Var}(\hat{P}_{\mathbf{m}_0})_{\Pi(|\Psi^{(3)}_r\rangle)}=\mathbf{g}_0^T\boldsymbol{\Omega}\boldsymbol{\gamma}_{\Pi(|\Psi^{(3)}_r\rangle)}\boldsymbol{\Omega}\mathbf{g}_0=\frac{3}{2}R^{(3)}_-,
\end{align}
and, together with Eq.~(\ref{eq:3modeX}), upon insertion into Eq.~(\ref{eq:chig}),
\begin{align}\label{eq:3msa}
\xi^{2}_{\mathbf{g}_0}(|\Psi^{(3)}_r\rangle)=\frac{1}{3}(1+2e^{-4r}).
\end{align}
Hence, $\xi^{2}_{\mathbf{g}_0}(|\Psi^{(3)}_r\rangle)<1$ for all $r>0$, which violates the separability condition~(\ref{eq:simplecovcrit}).

The violation of the separability condition is also observed for the other eigenvectors $\mathbf{g}_1$ and $\mathbf{g}_2$, however, they produce smaller values of $\xi^{-2}_{\mathbf{g}_0}$. Specifically they lead to $\xi^{2}_{\mathbf{g}_1}(|\Psi^{(3)}_r\rangle)=\xi^{2}_{\mathbf{g}_2}(|\Psi^{(3)}_r\rangle)=(2 + e^{-4 r})/3$. The reason for this is that the anti-squeezing along the respective maximally non-commuting direction is not in all cases reduced with equal effectiveness by the removal of correlations. In other words, $\mathbf{g}^T\boldsymbol{\Omega}^T\boldsymbol{\gamma}_{\Pi(\hat{\rho})}\boldsymbol{\Omega}\mathbf{g}$ is not necessarily small when $\mathbf{g}^T\boldsymbol{\gamma}_{\hat{\rho}}\mathbf{g}$ is.

\subsubsection{Non-Gaussian entanglement from higher-order squeezing}\label{sec:nlsq}
Squeezing is a second-order process that is able to generate entangled Gaussian states, such as the two examples discussed above. In order to illustrate how the squeezing coefficient can be adapted to non-Gaussian states, we consider a fourth-order squeezing evolution as a direct extension  of the (conventional) second-order squeezing, discussed in Sec.~\ref{sec:2ms}. Specifically, we consider states of the form
\begin{align}\label{eq:4thordersqz}
|\psi_r\rangle=\hat{X}_{12}[r]|0,0\rangle,
\end{align}
where $\hat{X}_{12}[\xi]=e^{\frac{1}{2}(\xi\hat{a}^{\dagger 2}_1\hat{a}^{\dagger 2}_2-\xi^*\hat{a}_1^2\hat{a}_2^2)}$. The fourth-order generator in $\hat{X}_{12}$ produces a squeezing effect which is not captured by the variances of linear observables, i.e., the covariance matrix. Hence, the criteria~(\ref{eq:MultiGiovannetti}) and~(\ref{eq:CVVariance}) are not able to detect the state's entanglement if the operators $\hat{A}_i$ and $\hat{B}_i$ are limited to quadratures. Consequently also Simon's PPT criterion for Gaussians states \cite{PhysRevLett.84.2726} cannot reveal this non-Gaussian entanglement.

We therefore extend the squeezing coefficient~(\ref{eq:chi2}) by employing nonlinear observables
\begin{align}\label{eq:2ndorderops}
\hat{D}(\boldsymbol{\mu})&=\mu_1(\hat{a}_1^2+\hat{a}_1^{\dagger 2})+\mu_2 i(\hat{a}_1^{\dagger 2}-\hat{a}_1^2)\notag\\&\quad+\mu_3(\hat{a}_2^2+\hat{a}_2^{\dagger 2})+\mu_4 i(\hat{a}_2^{\dagger 2}-\hat{a}_2^2),
\end{align}
characterized by a vector $\boldsymbol{\mu}=(\mu_1,\mu_2,\mu_3,\mu_4)\in \mathbb{R}^4$. From Eq.~(\ref{eq:CVVariance}), we find that the non-Gaussian squeezing coefficient
\begin{align}\label{eq:chi}
\chi_{\boldsymbol{\mu},\boldsymbol{\nu}}(\hat{\rho})=  \frac{4 \mathrm{Var}(\hat{D}(\boldsymbol{\mu}))_{\hat{\rho}} \mathrm{Var}(\hat{D}(\boldsymbol{\nu}))_{\Pi(\hat{\rho})}}{\left|\langle[\hat{D}(\boldsymbol{\mu}),\hat{D}(\boldsymbol{\nu})]\rangle_{\hat{\rho}}\right|^2}
\end{align}
must satisfy
\begin{align}\label{eq:entcrit}
\chi_{\boldsymbol{\mu},\boldsymbol{\nu}}(\hat{\rho}_{\mathrm{sep}})\geq 1 \quad \forall \boldsymbol{\mu},\boldsymbol{\nu}
\end{align}
for all separable states $\hat{\rho}_{\mathrm{sep}}$.

\begin{figure}[tb]
\centering
\includegraphics[width=.4\textwidth]{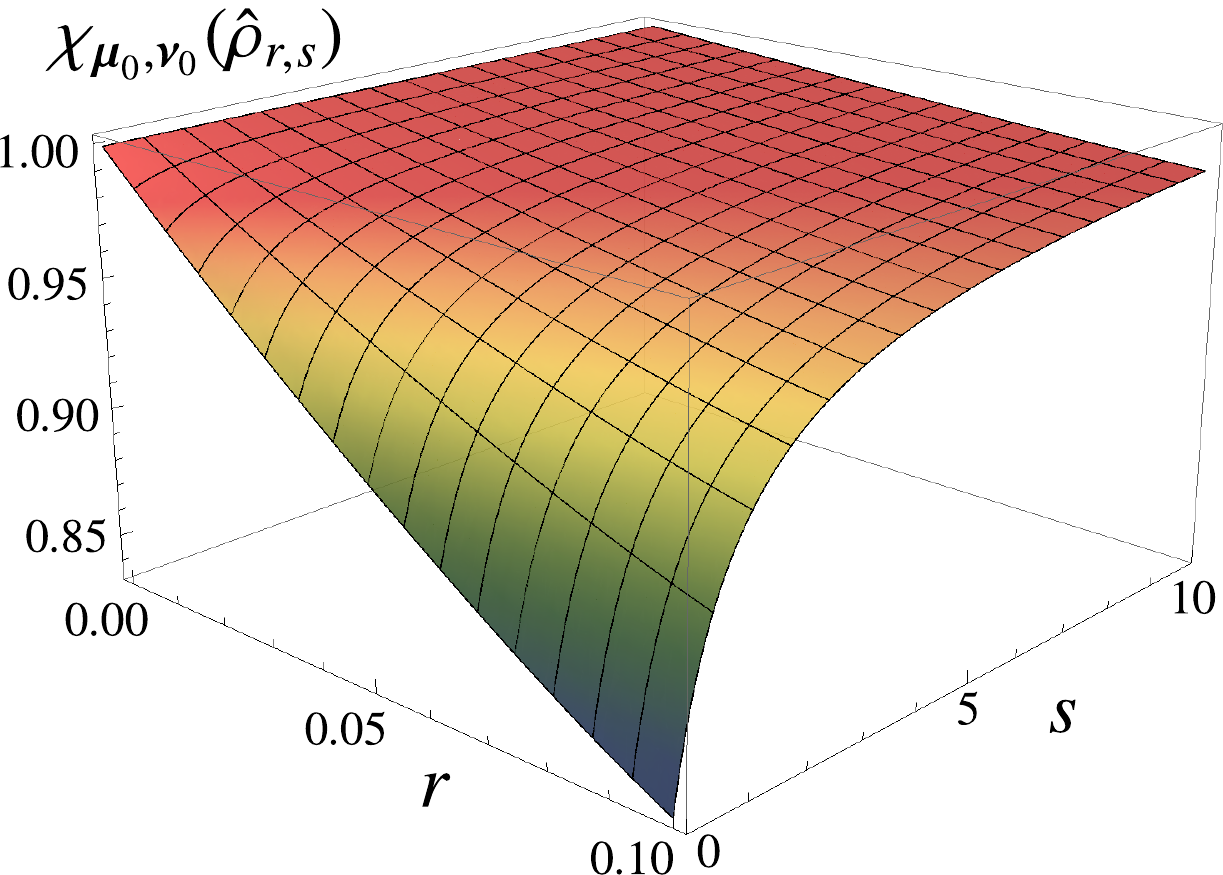}
\caption{Non-Gaussian squeezing coefficient $\chi_{\boldsymbol{\mu}_0,\boldsymbol{\nu}_0}(\hat{\rho}_{r,s})$, Eq.~(\ref{eq:chi}), for the fourth-order two-mode squeezed vacuum state, mixed with the vacuum, Eq.~(\ref{eq:rhorg}). The state's entanglement is revealed by the higher-order condition $\chi_{\boldsymbol{\mu}_0,\boldsymbol{\nu}_0}(\hat{\rho}_{r,s})<1$ [see Eq.~(\ref{eq:entcrit})] as long as $r>0$. Gaussian criteria which involve only the covariance matrix are not able to recognize the state as entangled.}
\label{fig:4thord}
\end{figure}

This novel criterion identifies the entanglement of the states
\begin{align}\label{eq:rhorg}
\hat{\rho}_{r,s}=\frac{1}{1+s}(|\psi_r\rangle\langle \psi_r|+s|0,0\rangle\langle 0,0|),
\end{align}
as is shown in Fig.~\ref{fig:4thord}. By mixing the state~(\ref{eq:4thordersqz}) incoherently with the vacuum, we further test the applicability of the criterion to mixed continuous-variable states. The optimal squeezing directions in the ``phase space'' of second-order variables, i.e., those that lead to the maximum value $\chi_{\boldsymbol{\mu}_0,\boldsymbol{\nu}_0}(\hat{\rho}_{r,s})$ are given by $\boldsymbol{\mu}_0=(c_1,c_2,-c_1,c_2)$ and $\boldsymbol{\nu}_0=\boldsymbol{\Omega}\boldsymbol{\mu}_0$ for all values of $r$ and $s$, where $c_1$ and $c_2$ are arbitrary real numbers. All these observations are in complete analogy to the results of Sec.~\ref{sec:2ms} once the conventional covariance matrix, generated by (linear) quadratures $\hat{a}_i$, is replaced by the nonlinear covariance matrix, generated by second-order quadratures $\hat{a}^{2}_i$.

With the help of the second-order observables~(\ref{eq:2ndorderops}), the entanglement of $\hat{\rho}_{r,s}$ can also be revealed by violation of condition~(\ref{eq:MultiGiovannetti}) with numerically optimal directions $\boldsymbol{\mu}_0'=\boldsymbol{\mu}_0$ and $\boldsymbol{\nu}_0'=(c_2,-c_1,-c_2,-c_1)$.

\section{Detecting entanglement in a fixed partition}\label{sec:multi}
So far our analysis was focused on detecting inseparability in a multipartite system, without information about which of the parties are entangled. 
It should be noticed that the variance on the right-hand side of the separability criterion~(\ref{eq:sepcriterion}) 
is calculated on the state $\Pi(\hat{\rho})=\bigotimes_{i=1}^N\hat{\rho}_i$, where $\hat{\rho}_i$ 
is the reduced density operator of $\hat{\rho}$ for subsystem $i$. The replacement of $\hat{\rho}$ by $\Pi(\hat{\rho})$ removes all correlations between the subsystems, see Sec.~\ref{sec:enhanced}. In a multipartite system this is just one of the possible ways to separating the full system into uncorrelated, possibly coarse-grained subsystems. The general criterion~(\ref{eq:sepcriterion}) can be extended to probe for entanglement in a specific partition of the system. 

To see this, let us introduce the generalized set of operations
\begin{align}\label{eq:kprojection}
\Pi_{\{\mathcal{A}_1,\dots,\mathcal{A}_M\}}(\hat{\rho})=\bigotimes_{l=1}^M\hat{\rho}_{\mathcal{A}_l},
\end{align}
where the $\{\mathcal{A}_1,\dots,\mathcal{A}_M\}$ label some coarse-grained partition of the $N$ subsytems and $\hat{\rho}_{\mathcal{A}_l}$ denotes the reduced density operator of $\hat{\rho}$ on the subsystems labeled by $\mathcal{A}_l$. We introduce the term $\{\mathcal{A}_1,\dots,\mathcal{A}_M\}$-separable for quantum states $\hat{\rho}_{\{\mathcal{A}_1,\dots,\mathcal{A}_M\}}$ that can be decomposed in the form
\begin{align}\label{eq:blocksep}
\hat{\rho}_{\{\mathcal{A}_1,\dots,\mathcal{A}_M\}}=\sum_{\gamma}p_{\gamma}\hat{\rho}^{(\gamma)}_{\mathcal{A}_1}\otimes\dots\otimes\hat{\rho}^{(\gamma)}_{\mathcal{A}_M},
\end{align}
where the $\hat{\rho}^{(\gamma)}_{\mathcal{A}_l}$ describe arbitrary quantum states on $\mathcal{A}_l$. We will now show that Eq.~(\ref{eq:sepcriterion}) can be extended to yield a condition for $\{\mathcal{A}_1,\dots,\mathcal{A}_M\}$-separability. Any $\{\mathcal{A}_1,\dots,\mathcal{A}_M\}$-separable state must satisfy
\begin{align}\label{eq:partitioncriterion}
F_Q[\hat{\rho}_{\{\mathcal{A}_1,\dots,\mathcal{A}_M\}},\hat{A}]&\leq 4\mathrm{Var}(\hat{A})_{\Pi_{\{\mathcal{A}_1,\dots,\mathcal{A}_M\}}(\hat{\rho})},
\end{align}
where $\hat{A}=\sum_{i=1}^N\hat{A}_i$. The proof is provided in Appendix~\ref{app:proof}.

The criterion~(\ref{eq:partitioncriterion}) allows us to devise witnesses for entanglement among specific, possibly coarse-grained subsets of the full system. Instead of removing all off-diagonal blocks from the covariance matrix, as was done in the case of full separability, one only puts certain off-diagonal blocks, as specified by the chosen partition, to zero. Different partitions then lead to different bounds on the Fisher information. In the extreme case with the trivial partition that contains all subsystems, we recover the upper bound $F_Q[\hat{\rho},\hat{A}]\leq 4\mathrm{Var}(\hat{A})_{\hat{\rho}}$, which holds for all quantum states $\hat{\rho}$ and is saturated by pure states \cite{PhysRevLett.72.3439}. In the opposite limit, when the partition separates all of the $N$ subsystems, we obtain Eq.~(\ref{eq:sepcriterion}).

Notice that for probing $\{\mathcal{A}_1,\dots,\mathcal{A}_M\}$-separability, one may also employ partially non-local operators, i.e., operators that can be non-local on the potentially entangled subsets and local between separable subsets. We will call the general class of such operators $\{\mathcal{A}_1,\dots,\mathcal{A}_M\}$-local. Let us consider the case of a pure state, for which $F_Q[|\Psi\rangle,\hat{A}]= 4\mathrm{Var}(\hat{A})_{|\Psi\rangle}$ holds. If condition~(\ref{eq:partitioncriterion}) is satisfied for all possible $\{\mathcal{A}_1,\dots,\mathcal{A}_M\}$-local operators, we can conclude that there are no correlations between the different subsystems \cite{Gessner2016}. Since this implies a product state, we observe that Eq.~(\ref{eq:partitioncriterion}) represents a necessary and sufficient condition for $\{\mathcal{A}_1,\dots,\mathcal{A}_M\}$-separability of pure states. More precisely, for each pure $\{\mathcal{A}_1,\dots,\mathcal{A}_M\}$-inseparable state, a $\{\mathcal{A}_1,\dots,\mathcal{A}_M\}$-local operator exists that violates Eq.~(\ref{eq:partitioncriterion}).

As these results are independent of the dimensions of the local Hilbert spaces $\mathcal{H}_i$, they apply to arbitrary multipartite systems of discrete, continuous and hybrid variables. Hence, a simple substitution of the product state $\Pi(\hat{\rho})$ by $\Pi_{\{\mathcal{A}_1,\dots,\mathcal{A}_M\}}(\hat{\rho})$ leads to a generalization of the criterion~(\ref{eq:sepcriterion}). The methods developed in Refs.~\cite{Gessner2016,ResolutionEnhanced} which follow from Eq.~(\ref{eq:sepcriterion}) can thus be easily extended to render them susceptible to entanglement in a specific partition. In particular, we may use this to modify the squeezing coefficients~(\ref{eq:genSC}) and variance-assisted Fisher densities~(\ref{eq:genFD}) accordingly, to detect $\{\mathcal{A}_1,\dots,\mathcal{A}_M\}$-separable states.

In the continuous-variable case, a generalization of the squeezing coefficient~(\ref{eq:chi2}) is found as
\begin{align}\label{eq:chi2mult}
&\quad\xi_{\{\mathcal{A}_1,\dots,\mathcal{A}_M\}}^2(\hat{\rho})\notag\\&:=\min_{\mathbf{g}}\frac{4(\mathbf{g}^T\boldsymbol{\Omega}^T\boldsymbol{\gamma}_{\Pi_{\{\mathcal{A}_1,\dots,\mathcal{A}_M\}}(\hat{\rho})}\boldsymbol{\Omega}\mathbf{g})(\mathbf{g}^T\boldsymbol{\gamma}_{\hat{\rho}}\mathbf{g})}{(\mathbf{g}^T\mathbf{g})^2},
\end{align}
and an observation of $\xi_{\{\mathcal{A}_1,\dots,\mathcal{A}_M\}}^{-2}(\hat{\rho})>1$ indicates inseparability in the $\{\mathcal{A}_1,\dots,\mathcal{A}_M\}$-partition.

Let us illustrate this with the aid of the three-mode squeezed state $|\Psi^{(3)}_r\rangle$, which was introduced in Sec.~\ref{sec:3ms}. To check for inseparability within the $\{(1,2),(3)\}$-partition, we employ the covariance matrix of the state $\Pi_{\{(1,2),(3)\}}$. This is obtained from the full covariance matrix~(\ref{eq:3msa}) by removing all correlations between the subsystems $(1,2)$ and $(3)$, while retaining correlations between $(1)$ and $(2)$. This yields
\begin{align}
&\qquad\boldsymbol{\gamma}_{\Pi_{\{(1,2),(3)\}}(|\Psi^{(3)}_r\rangle)}\notag\\&=\frac{1}{2}\begin{pmatrix} 
R_+^{(3)} & 0 & S^{(3)} & 0 & 0 & 0
\\ 0 & R_-^{(3)} & 0 & -S^{(3)} & 0 & 0
\\ S^{(3)} & 0 & R_+^{(3)} & 0 & 0 & 0
\\ 0 & -S^{(3)} & 0 & R_-^{(3)} & 0 & 0
\\ 0 & 0 & 0 & 0 & R^{(3)}_+ & 0
\\ 0 & 0 & 0 & 0 & 0 & R_-^{(3)}
\end{pmatrix}.
\end{align}
The phase-space direction $\mathbf{g}_0$ leads to
\begin{align}
\xi_{\{(1,2),(3)\}}^{2}(|\Psi^{(3)}_r\rangle)=\frac{1}{9}(5 + 4 e^{-4 r}),
\end{align}
which reveals inseparability in the $\{(1,2),(3)\}$-partition for all $r>0$. We obtain the same result for the partitions $\{(1),(2,3)\}$ and $\{(1,3),(2)\}$.

Let us remark, however, that in general, entanglement in all the possible partitions does not necessarily imply genuine multipartite entanglement \cite{PhysRevA.65.012107}. A state of the form $\hat{\rho}=p_1\hat{\rho}_{\{1,2\}}\otimes\hat{\rho}_{3}+p_2\hat{\rho}_1\otimes\hat{\rho}_{\{2,3\}}+p_3\hat{\rho}_{\{1,3\}}\otimes\hat{\rho}_{2}$ with $0<p_1,p_2,p_3<1$ and entangled states $\hat{\rho}_{\{1,2\}}$, $\hat{\rho}_{\{2,3\}}$ and $\hat{\rho}_{\{1,3\}}$ is not separable in any of the partitions $\{(1,2),(3)\}$, $\{(1),(2,3)\}$, or $\{(1,3),(2)\}$. Yet, it only contains bipartite entanglement.

\section{Conclusions}
We introduced a multi-mode squeezing coefficient to detect entanglement in continuous-variable systems. This coefficient is based on squeezing of a collective observable in the $2N$-dimensional phase space. The entanglement of two- and three-mode squeezed states is successfully revealed. Interestingly, in both examples the inverse squared squeezing coefficient coincides with the number of squeezed modes in the limit of strong squeezing (see Fig.~\ref{fig:CVMMS}). Generalizations to higher-order phase space variables further allow for the detection of entanglement in non-Gaussian states.

The multi-mode squeezing coefficient can be directly measured in continuous-variable photonic cluster states, where the full covariance matrix can be extracted \cite{Roslund2014,PhysRevA.89.053828,PhysRevA.91.032314,PhysRevLett.114.050501}. Homodyne measurement techniques can also be realized with systems of cold atoms \cite{HamleyNATPHYS2012, Peise2015, GrossNATURE2011}. In the future, cold bosonic atoms in optical lattices under quantum gas microscopes may provide an additional platform for continuous-variable entanglement, where local access to individual modes is available \cite{Bak2009,She2010}. Similarly, also phonons in an ion trap may provide a locally controllable continuous-variable platform for quantum information \cite{PhysRevA.77.040302,Abdelrahman}.

There are several ways to improve the performance of entanglement detection by modifying the definition of the multi-mode squeezing coefficient~(\ref{eq:chi2}). As was illustrated in Sec.~\ref{sec:nlsq}, the inclusion of local observables that are nonlinear in $\hat{a}_i$ and $\hat{a}_i^{\dagger}$, can yield stronger criteria. Furthermore, only the variance that is measured without correlations needs to pertain to a sum of local operators. Employing Eq.~(\ref{eq:CVVariance}) with a more general choice of $\hat{B}(\boldsymbol{\beta})$ is further expected to provide tighter bounds for the Fisher information. The strongest criteria will be obtained when the Fisher information is measured directly \cite{Gessner2016}. Based on the intuition obtained from the study of spin systems \cite{PhysRevLett.102.100401,Strobel424,LucaRMP,ResolutionEnhanced}, this is expected to be particularly important to detect entanglement in 
non-Gaussian states. However, since the properties of the (quantum) Fisher information are largely unexplored in the context of continuous-variable systems, a detailed understanding of the relationship among the entanglement criteria discussed here and in the literature \cite{PhysRevA.67.022320,PhysRevA.67.052315,PhysRevA.90.052321,PhysRevA.90.062337,PhysRevA.92.042328,PhysRevLett.84.2722,PhysRevLett.84.2726,PhysRevLett.114.050501,PhysRevLett.117.140504} is presently still lacking. For instance, it is not known whether the squeezing criterion is equivalent to the PPT condition~\cite{PhysRevA.67.022320,PhysRevLett.84.2722,PhysRevLett.84.2726} in the case of quadrature observables.

We have extended our techniques beyond the detection of full inseparability to witness entanglement in a particular partition of the multipartite system. A simple generalization of the squeezing coefficient presented here (and similarly of related methods developed in Refs.~\cite{Gessner2016,ResolutionEnhanced}), provides an additional tool to characterize the entanglement structure in a multipartite system on a microscopic level.

Finally, the approach presented here can be combined with the concept of generalized spin squeezing which was developed in~\cite{ResolutionEnhanced} to define a similar squeezing parameter and entanglement criterion for hybrid combinations of discrete- and continuous-variable systems \cite{Jeong2014,Morin2014}.

\acknowledgments
M.G. acknowledges support from the Alexander von Humboldt foundation. We thank G. Ferrini for useful discussions.

\appendix

\section{Extension of the entanglement criterion by Giovannetti et al. \cite{PhysRevA.67.022320} to multi-mode systems}\label{app:giovannettimulti}
Here we show that the proof presented for bipartite systems in \cite{PhysRevA.67.022320} can be straight-forwardly extended to multipartite systems, leading to Eq.~(\ref{eq:MultiGiovannetti}). First notice that for a separable state~(\ref{eq:fullysep}), the variance of a sum of local operators $\hat{A}(\boldsymbol{\alpha})=\sum_{i=1}^N\alpha_i\hat{A}_i$ is bounded by
\begin{align}
&\quad\mathrm{Var}(\hat{A}(\boldsymbol{\alpha}))_{\hat{\rho}_{\mathrm{sep}}}\notag\\&\geq \sum_{\gamma}p_{\gamma}\mathrm{Var}(\hat{A}(\boldsymbol{\alpha}))_{\hat{\rho}^{(\gamma)}_1\otimes\cdots\otimes\hat{\rho}^{(\gamma)}_N}\notag\\
&= \sum_{\gamma}p_{\gamma}\sum_{i,j=1}^N\alpha_i\alpha_j\mathrm{Cov}(\hat{A}_i,\hat{A}_j)_{\hat{\rho}^{(\gamma)}_1\otimes\cdots\otimes\hat{\rho}^{(\gamma)}_N}\notag\\
&= \sum_{\gamma}p_{\gamma}\sum_{i=1}^N\alpha_i^2\mathrm{Var}(\hat{A}_i)_{\hat{\rho}^{(\gamma)}_i}.
\end{align}
Analogously, we find $\mathrm{Var}(\hat{B}(\boldsymbol{\beta}))_{\hat{\rho}_{\mathrm{sep}}}\geq \sum_{\gamma}p_{\gamma}\sum_{i=1}^N\beta_i^2\mathrm{Var}(\hat{B}_i)_{\hat{\rho}^{(\gamma)}_i}$. The weighted sum of these two variances with $w_1,w_2\geq 0$ is bounded by
\begin{align}
&\quad w_1\mathrm{Var}(\hat{A}(\boldsymbol{\alpha}))_{\hat{\rho}_{\mathrm{sep}}}+w_2\mathrm{Var}(\hat{B}(\boldsymbol{\beta}))_{\hat{\rho}_{\mathrm{sep}}}\notag\\
&\geq \sum_{\gamma}p_{\gamma}\sum_{i=1}^N\left[w_1\alpha_i^2\mathrm{Var}(\hat{A}_i)_{\hat{\rho}^{(\gamma)}_i}+w_2\beta_i^2\mathrm{Var}(\hat{B}_i)_{\hat{\rho}^{(\gamma)}_i}\right]\notag\\
&\geq \sum_{\gamma}p_{\gamma}\sum_{i=1}^N\Bigg[w_1\alpha_i^2\mathrm{Var}(\hat{A}_i)_{\hat{\rho}^{(\gamma)}_i}\notag\\&\hspace{2.2cm}+w_2\beta_i^2\frac{\left|\langle [\hat{A}_i,\hat{B}_i]\rangle_{\hat{\rho}^{(\gamma)}_i}\right|^2}{4\mathrm{Var}(\hat{A}_i)_{\hat{\rho}^{(\gamma)}_i}}\Bigg]\notag\\
&\geq \sum_{\gamma}p_{\gamma}\sum_{i=1}^N\sqrt{w_1w_2}\left|\alpha_i\beta_i\langle [\hat{A}_i,\hat{B}_i]\rangle_{\hat{\rho}^{(\gamma)}_i}\right|,
\end{align}
where we used the Heisenberg-Robertson uncertainty relation,
\begin{align}
\mathrm{Var}(\hat{B}_i)_{\hat{\rho}^{(\gamma)}_i}\geq \frac{\left|\langle [\hat{A}_i,\hat{B}_i]\rangle_{\hat{\rho}^{(\gamma)}_i}\right|^2}{4\mathrm{Var}(\hat{A}_i)_{\hat{\rho}^{(\gamma)}_i}},
\end{align}
and the fact that the function $f(x)=c_1x+c_2/x$ over $x>0$ takes its minimum value at $2\sqrt{c_1c_2}$ \cite{PhysRevA.67.022320}. Next we use that for arbitrary operators $O$ and $\hat{\rho}_i=\sum_{\gamma}p_{\gamma}\hat{\rho}^{(\gamma)}_i$, we have $|\langle O\rangle_{\hat{\rho}_i}|=|\sum_{\gamma}p_{\gamma}\langle O\rangle_{\hat{\rho}^{(\gamma)}_i}|\leq \sum_{\gamma}p_{\gamma}|\langle O\rangle_{\hat{\rho}^{(\gamma)}_i}|$. Since the operators $\hat{A}_i$ and $\hat{B}_i$ act locally on $\mathcal{H}_i$, we may substitute $\langle [\hat{A}_i,\hat{B}_i]\rangle_{\hat{\rho}_i}=\langle [\hat{A}_i,\hat{B}_i]\rangle_{\hat{\rho}_{\mathrm{sep}}}$. We obtain
\begin{align}\label{eq:giovineq}
&\quad w_1\mathrm{Var}(\hat{A}(\boldsymbol{\alpha}))_{\hat{\rho}_{\mathrm{sep}}}+w_2\mathrm{Var}(\hat{B}(\boldsymbol{\beta}))_{\hat{\rho}_{\mathrm{sep}}}\notag\\
&\geq \sqrt{w_1w_2}\sum_{i=1}^N\left|\alpha_i\beta_i\langle [\hat{A}_i,\hat{B}_i]\rangle_{\hat{\rho}_{\mathrm{sep}}}\right|,
\end{align}
Maximizing the bound~(\ref{eq:giovineq}) over $\sqrt{w_1/w_2}$ \cite{PhysRevA.67.022320} finally yields
\begin{align}
&\quad\mathrm{Var}(\hat{A}(\boldsymbol{\alpha}))_{\hat{\rho}_{\mathrm{sep}}}\mathrm{Var}(\hat{B}(\boldsymbol{\beta}))_{\hat{\rho}_{\mathrm{sep}}}\notag\\&\geq\frac{\left(\sum_{i=1}^N\left|\alpha_i\beta_i\langle[\hat{A}_i,\hat{B}_i]\rangle_{\hat{\rho}_{\mathrm{sep}}}\right|\right)^2}{4}\notag,
\end{align}
which is the desired result, Eq.~(\ref{eq:MultiGiovannetti}).

\section{Proof of Eq.~(\ref{eq:partitioncriterion})}\label{app:proof}
In order to prove Eq.~(\ref{eq:partitioncriterion}), we follow the proof presented for the main result in Ref.~\cite{Gessner2016}. We obtain
\begin{align}
&F_Q\bigg[\hat{\rho}_{\{\mathcal{A}_1,\dots,\mathcal{A}_M\}},\sum_{i=1}^N\hat{A}_i\bigg]\notag\\&\leq \sum_{\gamma}p_{\gamma}F_Q\left[\hat{\rho}^{(\gamma)}_{\mathcal{A}_1}\otimes\dots\otimes\hat{\rho}^{(\gamma)}_{\mathcal{A}_M},\sum_{i=1}^N\hat{A}_i\right]\notag\\
&\leq 4\sum_{\gamma}p_{\gamma}\mathrm{Var}(\sum_{i=1}^N\hat{A}_i)_{\hat{\rho}^{(\gamma)}_{\mathcal{A}_1}\otimes\dots\otimes\hat{\rho}^{(\gamma)}_{\mathcal{A}_M}}\notag\\
&= 4\sum_{\gamma}p_{\gamma}\sum_{l=1}^M\mathrm{Var}(\sum_{i\in\mathcal{A}_l}\hat{A}_i)_{\hat{\rho}^{(\gamma)}_{\mathcal{A}_l}}\notag\\
&\leq 4\sum_{l=1}^M\mathrm{Var}(\sum_{i\in\mathcal{A}_l}\hat{A}_i)_{\hat{\rho}_{\mathcal{A}_l}}\notag\\
&=4\mathrm{Var}(\sum_{i=1}^N\hat{A}_i)_{\Pi_{\{\mathcal{A}_1,\dots,\mathcal{A}_M\}}(\hat{\rho})},
\end{align}
where we used the convexity and additivity properties of the Fisher information \cite{Varenna}, the concavity of the variance, as well as the relation $F_Q[\hat{\rho},\hat{H}]\leq 4\mathrm{Var}(\hat{H})_{\hat{\rho}}$ \cite{PhysRevLett.72.3439} and $\hat{\rho}_{\mathcal{A}_l}=\sum_{\gamma}p_{\gamma}\hat{\rho}^{(\gamma)}_{\mathcal{A}_l}$. 

Notice that the result can be extended to a more general class of $\{\mathcal{A}_1,\dots,\mathcal{A}_M\}$-local generators: Instead of the fully local operator $\sum_{i=1}^N\hat{A}_i$ considered here, one may employ operators of the form $\sum_{l=1}^M\hat{O}_{\mathcal{A}_l}$, where $\hat{O}_{\mathcal{A}_l}$ is an \textit{arbitrary} operator on the subsets contained in $\mathcal{A}_l$.


%

\end{document}